\documentclass[format=acmsmall, review=false]{acmart}
\usepackage{acm-ec-26}

\usepackage{booktabs}
\usepackage[ruled]{algorithm2e}

\usepackage{caption}
\usepackage{tikz,csquotes}
\usepackage[inline]{enumitem}
\usepackage{amsmath,amsthm}
\usepackage{todonotes}
\usepackage{mathtools,wrapfig}
\usepackage{hyperref,cleveref}

\theoremstyle{remark}
\newtheorem{assumption}{Assumption}
\newtheorem{remark}{Remark}
\newtheorem{observation}{Observation}
\newcommand{\PnL}{\mathrm{PnL}}
\DeclareMathOperator{\OptMM}{Opt_{MM}}

\setcitestyle{acmnumeric}

\title[Performative Market Making]{Performative Market Making}

\author{Charalampos Kleitsikas}
\orcid{0009-0000-9291-1976}
\email{charalampos.kleitsikas@kcl.ac.uk}
\affiliation{%
  \institution{King's College London}
  \city{London}
  \country{United Kingdom}
}

\author{Stefanos Leonardos}
\orcid{0000-0002-1498-1490}
\email{stefanos.leonardos@kcl.ac.uk}
\affiliation{%
  \institution{King's College London}
  \city{London}
  \country{United Kingdom}
}

\author{Carmine Ventre}
\orcid{0000-0003-1464-1215}
\email{carmine.ventre@kcl.ac.uk}
\affiliation{%
  \institution{King's College London}
  \city{London}
  \country{United Kingdom}
}
\begin{abstract}
Economic models adopted by trading agents do not merely analyse markets, but actively shape them. This effect, known as \emph{performativity}, arises when strategies follow established economic theories, resulting in a collective alteration of market dynamics by creating self-fulfilling prophecies. Although discussed in the literature on economic sociology, this deeply rooted phenomenon lacks mathematical formulation in financial markets. Our paper closes this gap by studying performativity in the context of market making in financial markets, by breaking down the canonical separation of diffusion processes between the description of the market environment and the agents' models. We do that by embedding the models in the process itself, creating a closed feedback loop, demonstrating how prices change towards greater conformity to the prevailing financial models used in the market. In this framework, we show with closed-form solutions how an agent should best respond to the current dominant strategies of other agents in the market and effectively exploit them while maintaining competitive actions and superior performance. We further prove that there is a performative stable state, where agents do not have an incentive in finding new strategies to which they deviate. Our work emphasises performativity as a naturally emerging phenomenon in markets. 
\end{abstract}

\begin{document}

\maketitle
\setcounter{tocdepth}{1} 
\tableofcontents
\newpage

\section{Introduction}\label{intro}

{Game theory has allowed us to develop intuitive and analytical results to formally study strategic behaviour of agents}. However, the environment in which the agents operate is decoupled from the strategies adopted by them. This assumption appears to be problematic in (financial) markets where the actions adopted by the agents (e.g., orders submitted) actively change the environment (e.g., the asset price) and ultimately the agents' payoffs. 
%
In fact, economic models and theories are so deeply ingrained in modern financial practice that they decisively shape the markets they are meant to predict. This phenomenon, known as \emph{performativity}, challenges the orthodox scientific view that separates an observed \enquote{world} from the \enquote{language} used to describe it. In its \enquote{pure}, also known as \enquote{Barnesian}, form, \emph{financial performativity} is defined as 
{\setlength{\leftmargini}{12pt}%
\begin{quote}
\emph{[the] practical use of an aspect of economics that makes the economic processes more like their depiction by economics} \cite{MacKenzie2006}.
\end{quote}}
\noindent Unlike most scientific domains, where theories seek only to describe external reality, the extensive application of financial models creates powerful feedback loops with the markets on which they act. While performativity is a well-documented concept in economic sociology and has recently been formalized in supervised learning, it lacks a quantitative formulation in the context of financial market making \cite{MacKenzie2025, perdomo21}. Our work closes this gap by embedding model-driven dynamics directly into the stochastic processes that govern price evolution.



\subsection*{Our Contributions}
We study performativity in the context of high-frequency market making, breaking the canonical separation between price diffusion and agents' models. In market making, agents submit two sided-orders (a buy and a sell order) for 
an asset, in the hope of making money from the spread. This is a core liquidity-provision strategy, widely adopted in financial markets, which keeps trading active and supports price discovery. In practice, these two sided-orders are mostly skewed and create distribution shifts that can reflect either the beliefs of the market makers on the future price movement or the endogenous dynamics created by how their strategies manage internal constraints, such as excessive inventory from filled orders. However, the strategy is inherently risky, as it exposes agents to inventory risk and model-induced losses, and may generate endogenous instabilities at the market level. Against this background, our contributions are as follows.

\begin{itemize}[left=0pt, topsep=3pt]
    \item \textbf{Mathematical Framework:} We propose a formal model where the asset mid-price $s_t$ is mean reverting to the reference process $r_t$, representing the valuation implied by dominant financial models. To the best of our knowledge, this is the first quantitative model for performativity in financial markets. 
    \item \textbf{Performative Best Response:} We derive closed-form solutions for a ``performativity-aware'' agent. We demonstrate how such an agent can infer and exploit the predictable price drift generated by performativity-unaware competitors (e.g., those using standard Avellaneda-Stoikov strategies \cite{avellaneda2008}, referred to as A\&S below) while maintaining superior performance.
    \item \textbf{Performative Stability:} We prove the existence of a \textit{performative stable state}---a fixed point where the strategies adopted by agents shift the price distribution in a way that leaves the model optimal.
    \item \textbf{Empirical Validation:} We complement our theoretical results with simulations demonstrating that performativity-aware strategies consistently outperform classical baselines as the performative sensitivity $\epsilon$ increases.
\end{itemize}

\noindent By placing financial models at the core of price formation, we highlight performativity as a naturally emerging phenomenon that fundamentally alters the game-theoretic landscape of modern markets.

\subsection*{Technical Overview}
To model performativity in a general financial setting, we proceed axiomatically by positing that the evolution of an asset's mid-price, denoted as $s_t$, is influenced by a reference process $r_t$, representing the valuation implied by prevailing strategies. This results in a stochastic process formulation:
\begin{equation} \label{eq1}
ds_t = \underbrace{\epsilon\left( r_t -  s_t \right)dt}_{\text{performative effect}}\,\,\,+\hspace{-5pt} \underbrace{\sigma\, dW_t,}_{\text{original process}}
\end{equation}
where $\epsilon > 0$ governs the intensity of the performative feedback loop. Unlike standard diffusion processes that assume a separation between market environment and agent models, this framework embeds model-driven valuations directly into price dynamics.\footnote{The structure of Equation~\eqref{eq1} resembles the general form of the Ornstein–Uhlenbeck process with time-varying drift, commonly used in interest rate modeling. However, our use differs both in motivation and interpretation: here, he mean-reverting term embodies the influence of prevailing financial models on price formation, not externally given interest rate target and mean-reversion speed.}

We instantiate this framework for market making by defining a formal optimization operator, $\OptMM(\cdot)$, which maps a perceived price process to an optimal quoting strategy under either linear or exponential utility functions. Our theoretical results involve solving the Hamilton-Jacobi-Bellman (HJB) equations associated with these induced price distributions. We characterize \textit{Performative Optimality} as a single-step re-optimisation against a fixed reference strategy $r^*$, and \textit{Performative Stability} as the fixed point $r_{PS}^* = \OptMM(r_{PS}^*)$ where the strategy remains invariant under repeated HJB optimisation.

\subsection*{Related Work} We review relevant literature across three axes. \par

\smallskip\noindent \emph{Performativity in finance.} The notion of performativity in social predictions was originally described in ~\cite{Merton1948} and ~\cite{Grunberg1954}, with ~\cite{Barnes1983} offering a broader conceptual treatment. MacKenzie’s analysis of the Chicago Board Options Exchange~\cite{MacKenzie2003} and his subsequent overview~\cite{MacKenzie2006} examined performativity in financial economics. In their recent review, ~\cite{MacKenzie2025} survey 6,741 works that mention performativity across various fields but find no quantitative frameworks in financial mathematics.\par
\smallskip\noindent \emph{Performativity in machine learning.}  
Inspired by the aforementioned works, the notion of performativity was introduced in the field of supervised learning by \cite{perdomo21}, triggering a body of work quantifying how predictions influence their targets~\cite{mendler20, perdomo21b, brown22a, Dunner22} and analysing the concepts of optimality and stability in such settings. Extensions include reinforcement-learning settings~\cite{mandal23} and multi-agent environments~\cite{piliouras22}. In Section~\ref{conclusions}, we outline how our framework complements these efforts as AI-driven models increasingly shape market behaviour.\par
\smallskip \noindent  \emph{Market Making.} The limit-order market-making problem was formalised and popularised by the work of \cite{avellaneda2008} which was inspired by an old paper on a dealer's optimal pricing, that of \cite{Ho1981}. The core idea was to use classical tools from stochastic optimal control theory and describe the problem with a model-based approach. That sparked a long line of literature ~\cite{fodra2012, Gu_ant_2012, cartea14, Cartea2015, cartea17} that developed a wide range of variations and extensions of the original model and provided analytical solutions. A collective presentation and a textbook-level treatment of these approaches are provided in ~\cite{cartea_book, guent_book}. Market making has also been examined with modern techniques, mainly that of Reinforcement Learning, e.g. ~\cite{lim2018}, with most of them keeping the model-based approach to formulate the dynamics of the environment ~\cite{spooner2020, mbt-gym, carmine1, carmine2}. The success of \citet{avellaneda2008} mathematical framework and model to describe fundamental properties of market making alongside its  widespread usage, makes it a natural use case to study performativity in high-frequency trading. \par

\section{Preliminaries: Market-Making Model} \label{preliminaries}

We consider a continuous-time market for a single traded asset over a finite horizon $[0,T]$, defined on a filtered probability space $(\Omega,\mathcal{F},(\mathcal{F}_t)_{t\in[0,T]},\mathbb{P})$ satisfying the usual conditions \cite{Oks98}.

\paragraph{Price Dynamics.} 
The mid-price denoted by $(s_t)_{t\in[0,T]}$ is the average between the best bid and the best ask. As common in the literature, we assume that it follows an It\^o diffusion
\begin{equation} \label{eq:price}
ds_t = b(t,s_t)\,dt + \sigma(t,s_t)\,dW_t,
\end{equation}
where $b(t,s_t)$ denotes the \textit{drift term}, and $(W_t)_t$ is a standard Brownian motion adapted to $(\mathcal{F}_t)_t$. When $b \equiv 0$, the price process is a martingale, consistent with the weak-form market efficiency hypothesis.

\paragraph{Market-Maker Quotes and Reservation Price.}

A market maker (MM) provides liquidity by continuously quoting an ask price $p^a_t$ and a bid price $p^b_t$, typically above and below the mid-price $s_t$.
At time $t$, the MM’s state is described by the tuple $(x_t,s_t,q_t)$,
where $x_t \in \mathbb{R}$ denotes cash holdings,
$s_t$ the mid-price,
and $q_t \in \mathbb{Z}$ the inventory position. The ask and bid premia are defined as
$\delta^a_t = p^a_t - s_t$ and $\delta^b_t = s_t - p^b_t$,
so that the quoted \emph{spread} is $\delta_t = \delta^a_t + \delta^b_t$.
The MM’s \emph{reservation price} is defined as the midpoint between the quotes,
\begin{equation}\label{eq:reservation}
r_t = \frac{p^a_t + p^b_t}{2}
    = s_t + \frac{\delta^a_t - \delta^b_t}{2}.
\end{equation}
The reservation price captures the MM’s internal valuation of the asset,
reflecting inventory considerations and directional beliefs.

\paragraph{Order Arrivals}
Market buy and sell orders arrive according to independent Poisson processes $(N^a_t)_t$ and $(N^b_t)_t$ with intensities
\begin{equation} \label{eq:intensities}
\lambda^a(\delta^a_t) = A e^{-k\delta^a_t}, \quad
\lambda^b(\delta^b_t) = A e^{-k\delta^b_t},
\end{equation}
where $A,k>0$ are market parameters capturing \emph{order frequency} and \textit{book depth}\footnote{Book depth is the amount of resting buy/sell liquidity (i.e., live orders) in the market.}, respectively, with tighter quotes leading to higher execution intensity.

\paragraph{State Variables and PnL}

The MM’s\textit{inventory} $q_t \in \mathbb{Z}$ and \textit{cash position} $x_t \in \mathbb{R}$ follow the dynamics
\[
dq_t = dN^b_t - dN^a_t,
\qquad
dx_t = (s_t+\delta^a_t)\,dN^a_t - (s_t-\delta^b_t)\,dN^b_t.
\]
Accordingly, the MM's wealth or \textit{profit and loss (PnL)} evolves as
\begin{equation}\label{eq:pnl}
    \PnL_t =x_t +q_ts_t.
\end{equation}

\paragraph{Objective}
The market maker selects predictable controls $(\delta^a_t,\delta^b_t)_{t\in[0,T]}$ to maximize expected utility of terminal wealth over a finite horizon $T$.
For a utility function $\phi:\mathbb{R}^3 \to \mathbb{R}$, the MM's \textit{value function} is defined as
\begin{equation}
H(t,x,s,q)
=
\sup_{\delta^a,\delta^b \in \mathcal{A}}
\mathbb{E}_{t,x,s,q}\!\left[\phi(x_T,s_T,q_T)\right],
\end{equation}
where $\mathcal{A}$ denotes the set of admissible controls, i.e., $\mathcal{F}$-predictable processes bounded from below. Recall that here the state variables $(x_t,s_t,q_t)$ denote the MM’s cash holdings, mid-price, and inventory at time $t$, respectively.

\begin{example} [Utility Specifications] We consider two standard utility specifications; \textit{linear} and \textit{exponential} utilities. For linear utility, $\phi(x,s,q) = x + q s$, the value function reduces to
\begin{equation*}
H(t,x,s,q)=
\sup_{\delta^a,\delta^b \in \mathcal{A}}
\mathbb{E}_{t,x,s,q}\!\left[x_T + q_T s_T\right],
\end{equation*}
corresponding to the maximization of expected terminal $\PnL$, where inventory is liquidated at the terminal price $s_T$. Exponential utility, $\phi(x,s,q) = -\exp\{-\gamma(x + q s)\}$, with $\gamma>0$, accounts for inventory risk, yielding the value function
\begin{equation*}
H(t,x,s,q)=
\sup_{\delta^a,\delta^b \in \mathcal{A}}
\mathbb{E}_{t,x,s,q}\!\left[-\exp\{-\gamma(x_T + q_T s_T)\}\right].   
\end{equation*}
This specification penalizes large inventory positions and induces risk-sensitive quoting behaviour.
\end{example}

\section{Performativity Framework}

We formalize performativity by allowing the evolution of an economic variable to depend on a reference model that is itself produced by agents operating in the market.

\subsection{Performative Price Dynamics}
In the context of financial markets, this reference represents the valuation implied by a dominant strategy or financial model and feeds back into price formation.

\begin{definition}[Performative Price Process]\label{def:ppp}
Let $(r_t)_{t\in[0,T]}$ be an adapted stochastic process representing a reference valuation generated by a prevailing economic model or strategy.
The mid-price $(s_t)_{t\in[0,T]}$ follows a \emph{performative price process} if it satisfies
\begin{equation} \label{eq:performative_price}
ds_t
=
\epsilon\bigl(r_t - s_t\bigr)\,dt
+
\sigma\,dW_t,
\end{equation}
where $\epsilon>0$ is the \emph{performativity parameter}, $\sigma>0$ the volatility, and $(W_t)_t$ a standard Brownian motion.
\end{definition}

Parameter $\epsilon$ governs the strength of the feedback from the reference model to the price process. When $\epsilon=0$, the price follows an exogenous martingale; as $\epsilon$ increases, the price increasingly conforms to the reference valuation. Equation~\eqref{eq:performative_price} introduces a feedback loop between prices and economic models. The reference process $r_t$ is not interpreted as a fundamental value, but as the valuation implied by agents’ strategies.
Performativity arises because widespread use of the same model causes prices to move toward the valuations that the model itself generates.

\paragraph{Solution of the Performative Price Process}

The performative price process in Definition~\ref{def:ppp} admits a closed-form solution.
Fix $t<T$ and assume that the reference process $(r_u)_{u\in[t,T]}$ is adapted and integrable.
Then the solution of equation \eqref{eq:performative_price} is given by
\begin{equation} \label{eq:perf_solution}
s_T
=
e^{-\epsilon (T-t)} s_t
+
\int_t^T \epsilon e^{-\epsilon (T-u)} r_u \, du
+
\sigma \int_t^T e^{-\epsilon (T-u)} dW_u .
\end{equation}
Consequently, $s_T$ is normally distributed with mean and variance
\[
\mathbb{E}[s_T \mid \mathcal{F}_t]
=
e^{-\epsilon (T-t)} s_t
+
\int_t^T \epsilon e^{-\epsilon (T-u)} r_u \, du,
\quad
\mathrm{Var}(s_T \mid \mathcal{F}_t)
=
\frac{\sigma^2}{2\epsilon}\bigl(1-e^{-2\epsilon(T-t)}\bigr).
\]
Equation~\eqref{eq:perf_solution} highlights the key economic effect of performativity.
The future price is a convex combination of the current price and an exponentially weighted average of the reference valuation process.
As $\epsilon$ increases, the influence of the reference model dominates both the mean and the variance of the price distribution.
In the limit of strong performativity, prices rapidly conform to the valuation implied by the prevailing strategy.

\begin{remark}[Martingale Property]
Unless $r_t \equiv s_t$, the performative price process is not a martingale.
The endogenous drift $\epsilon(r_t-s_t)$ introduces predictable components and a strategy-dependent directional bias into price dynamics. This creates a systematic deviation from market efficiency that may be exploited or arbitraged by \emph{performativity-aware} agents, i.e. agents who account for performative effects.
\end{remark}

\subsection{Performative Optimality}

Performativity alters the optimization problem faced by agents because the price distribution to which they respond is itself shaped by the strategies deployed in the market. As a result, optimality must be defined relative to the endogenous price process induced by a reference model.\par

Consider a market maker (MM) whose optimal strategy in a non-performative setting is given by a reservation price process $r^*$, obtained by solving the standard optimization problem under an exogenous price process.
Equivalently, $r^*$ summarizes the optimal quoting strategy $(\delta^{a,*},\delta^{b,*})$. In a performative setting, systematic deployment of the strategy $r^*$—for instance due to strategy crowding or monopolistic influence—feeds back into price formation.
As described above, this induces a performative price process of the form
\[
ds_t = \epsilon\bigl(r^* - s_t\bigr)\,dt + \sigma\,dW_t,
\]
where, for simplicity, the time dependence of $r^*$ is suppressed.
An agent who is aware of this feedback no longer faces the original optimization problem, but instead optimizes against the price dynamics induced by the deployed strategy $r^*$.
We refer to the resulting optimal strategy as the \emph{performative optimum}.

\begin{definition}[Performative Optimality]
Let $\OptMM(r)$ denote the MM’s optimization problem when the quoting strategy $r$ induces the price dynamics faced by the agent.
Given a reference strategy $r^*$ inducing the performative price process above, a strategy $r^*_{\mathrm{PO}}$ is \emph{performatively optimal} if it satisfies
\begin{equation}\label{eq:po}
    r^*_{\mathrm{PO}} = \OptMM(r^*).
\end{equation}
\end{definition}

In general, the performatively optimal strategy $r^*_{\mathrm{PO}}$ need not coincide with the original strategy $r^*$.
Economically, $r^*_{\mathrm{PO}}$ can be interpreted as the strategy adopted either by agents who infer and exploit the performative impact of $r^*$ on prices, or by dominant agents who internalize the effect their own strategy has on price formation and adjust accordingly.

Performative optimality therefore differs from classical optimality: the strategy is not evaluated against an exogenous environment, but against an environment endogenously shaped by the reference model.
When $\epsilon=0$, the price process is exogenous and independent of the reference strategy, and performative optimality reduces to the standard notion of optimality in stochastic control. \par
\noindent \textbf{Game-theoretic interpretation.} The typical game-theoretic modelling of the game played amongst different market makers would assume the exogenous price diffusion price \eqref{eq:price}. In the presence of market forces, the best responses calculated within this model would \emph{not} maximise the agents' payoffs as the right price diffusion model would in fact be \eqref{eq:performative_price}. Our approach highlights that the typical modelling ignores these feedback loops between strategies adopted and the game environment. 

\subsection{Performative Stability}

If $r^*_{\mathrm{PO}} \neq r^*$, systematic deployment of $r^*_{\mathrm{PO}}$ induces a new price process, which in turn alters the optimization problem faced by agents.
This iterative re-optimization dynamic motivates the performative stability notion introduced next. In particular, if the performatively optimal strategy $r^*_{\mathrm{PO}}$ differs from the original strategy $r^*$, its systematic deployment will in turn induce a new performative price process,
\begin{equation}
ds_t = \epsilon\bigl(r^*_{\mathrm{PO}} - s_t\bigr)\,dt + \sigma\,dW_t.
\end{equation}
Agents optimizing against this new environment will generally adjust their strategies again, giving rise to a sequence of re-optimizations and an evolving sequence of induced price dynamics.

From both a market-efficiency perspective and the perspective of agents whose strategies shape prices, it is natural to ask whether this re-optimization process converges.
In particular, one may ask whether there exist strategies that remain optimal even after accounting for their own performative effects.

\begin{definition}[Performative Stability]
A strategy $r^*_{\mathrm{PS}}$ is \emph{performatively stable} if it is performatively optimal with respect to the price process it induces.
That is, when prices evolve according to
\begin{equation*}
ds_t = \epsilon\bigl(r^*_{\mathrm{PS}} - s_t\bigr)\,dt + \sigma\,dW_t,
\end{equation*}
the strategy $r^*_{\mathrm{PS}}$ satisfies
\begin{equation}
r^*_{\mathrm{PS}} = \OptMM(r^*_{\mathrm{PS}}).
\end{equation}
\end{definition}

A performatively stable strategy therefore constitutes a fixed point of the re-optimization process: it is an economic model whose deployment shifts the price distribution in a way that leaves the model optimal.
As we show in the subsequent sections, performatively optimal strategies obtained through re-optimization do not generally coincide with performatively stable strategies. 

\noindent \textbf{Game-theoretic interpretation.} The step from performatively optimal strategies to performatively stable states is similar to that of any economic equilibrium concept. In these strategy-dependent environments, it is not sufficient to calculate the players' best responses since one is also interested in the equilibrium, the fixed point of these dynamics.  

\begin{remark}[Market-Making Optimisation under a Performative Price Process]

We briefly recall the standard market-making optimisation problem in order to make explicit the optimisation operator $\OptMM(\cdot)$ used throughout Section~3. Consider a market maker with state variables $(x_t,s_t,q_t)$ and utility function $\phi(x,s,q)$ as defined in Section~2.
Given a price process $(s_t)_{t\in[0,T]}$, the market maker chooses predictable quoting controls $(\delta^a_t,\delta^b_t)$ to maximise expected terminal utility.
The corresponding value function satisfies the Hamilton--Jacobi--Bellman (HJB) equation
\begin{align} \label{eq:hjb_generic}
(\partial_t + \mathcal{L})H
&+ \sup_{\delta^a} \lambda^a(\delta^a)\!\left[H(t,s,q-1,x+s+\delta^a) - H(t,s,q,x)\right] \nonumber\\
&+ \sup_{\delta^b} \lambda^b(\delta^b)\!\left[H(t,s,q+1,x-(s-\delta^b)) - H(t,s,q,x)\right]
= 0,
\end{align}
with terminal condition $H(T,x,s,q)=\phi(x,s,q)$.
Here, $\mathcal{L}$ denotes the infinitesimal generator associated with the price process $(s_t)$. Under standard regularity conditions, the HJB admits a unique solution and yields optimal controls that can be expressed in terms of an optimal reservation price process, denoted by $r^*$. We abstract this dependence via the optimisation operator $\OptMM(\cdot)$.
\end{remark}

The subsequent sections instantiate these concepts for concrete market-making models, characterize performative optima and stable strategies explicitly, and demonstrate how performativity-aware strategies can outperform classical optimal strategies in markets with endogenous price feedback.

\section{Performative Market Making under Endogenous Beliefs} \label{beliefs}

We instantiate the performativity framework in a simple and transparent setting in which a market maker’s performative influence arises from her endogenous beliefs about future price dynamics.
The purpose of this section is threefold:
(i) to illustrate how performative optimality and stability can be computed in closed form;
(ii) to show that performativity-aware strategies can outperform classical optimal strategies;
and, (iii) to motivate the role of performative stability as a coherent long-run solution concept.

\subsection{Baseline: Optimal Market Making under Directional Beliefs}

We begin with a benchmark setting in which a market maker (MM) holds directional beliefs about the asset’s future price evolution.
Specifically, suppose that at time $t$ the MM believes that the mid-price follows an arithmetic Brownian motion
\begin{equation} \label{eq:belief_abm}
ds_u = \mu\,du + \sigma\,dW_u, \qquad u \in [t,T],
\end{equation}
where $\mu \in \mathbb{R}$ represents the perceived average return and $\sigma>0$ the volatility.
We assume linear utility, $\phi(x,s,q)=x+qs$.

\begin{lemma}[Optimal Reservation Price under Directional Beliefs]
\label{lem:baseline_opt}
Under the price dynamics~\eqref{eq:belief_abm} and linear utility, the optimal reservation price satisfies
\[
r^*(t,s) = \mathbb{E}_{t,s}[S_T] = s + \mu(T-t),
\]
and the optimal bid and ask premia are given by
\[
\delta_*^{a,b} = \frac{1}{k} \pm \mu(T-t),
\qquad
\psi^* = \delta_*^a + \delta_*^b = \frac{2}{k}.
\]
where $k>0$ is the order-book depth parameter that controls the sensitivity of order arrival rates to quoted prices, cf. Equation~ \eqref{eq:intensities}.
\end{lemma}

\begin{proof}
Under linear utility $\phi(x,s,q)=x+qs$, the market maker is risk-neutral with respect to price fluctuations and inventory holdings.
In this case, the optimal reservation price coincides with the conditional expectation of the terminal price, as is standard in market-making models with linear utility. Solving the stochastic differential equation
\eqref{eq:belief_abm}, i.e., $ds_u = \mu\,du + \sigma\,dW_u$, $u\in[t,T]$, by direct integration yields
\[
S_T = s + \int_t^T \mu\,du + \sigma(W_T-W_t)
= s + \mu(T-t) + \sigma(W_T-W_t).
\]
Taking conditional expectations with respect to $\mathcal{F}_t$ gives $
\mathbb{E}_{t,s}[S_T] = s + \mu(T-t)$. The expressions for the optimal bid and ask premia follow from the solution of the market-making optimisation problem under Poisson order arrivals with exponential intensities, which yields symmetric optimal spreads and a reservation price equal to the conditional expectation of the terminal price (see, e.g., \cite{fodra2012, cartea_book}).
\end{proof}

The MM therefore quotes symmetrically around $r^*$ rather than the current mid-price $s$.
When $\mu>0$, this induces a directional skew: the bid quote is placed closer to $s$ and the ask quote further away, reflecting the MM’s belief that the asset will appreciate.

\subsection{Performative Optimality under Endogenous Beliefs}

We now introduce performativity.
Suppose that the MM’s quoting strategy, summarised by the reservation price $r^*$, is deployed systematically and feeds back into price formation.
Under the performative price process, the mid-price evolves according to
\begin{equation} \label{eq:belief_perf}
ds_u = \epsilon\bigl(r^*(t,s) - s_u\bigr)\,du + \sigma\,dW_u
= \epsilon \mu (T-u)\,du + \sigma\,dW_u,
\qquad u\in[t,T].
\end{equation}

A performativity-aware MM recognises that~\eqref{eq:belief_perf}, rather than~\eqref{eq:belief_abm}, governs price evolution and optimises accordingly.

\begin{proposition}[Performatively Optimal Strategy]
\label{prop:po_beliefs}
Under the performative price process~\eqref{eq:belief_perf} and linear utility, the performatively optimal reservation price and corresponding optimal premia are
\[
r^*_{\mathrm{PO}}(t,s)
=
\mathbb{E}_{t,s}[S_T]
=
s + \epsilon \mu \frac{(T-t)^2}{2}, \qquad\delta^{a,b}_{\mathrm{PO}} = \frac{1}{k} \pm \epsilon \mu \frac{(T-t)^2}{2}.
\]
\end{proposition}

\begin{proof}
Equation~\eqref{eq:belief_perf} is again an arithmetic Brownian motion with time-dependent drift.
Integrating over $[t,T]$ yields
\[
S_T = s + \int_t^T \epsilon\mu(T-u)\,du + \sigma(W_T-W_t)
= s + \epsilon \mu \frac{(T-t)^2}{2} + \sigma(W_T-W_t).
\]
Since the MM has linear utility, the performatively optimal reservation price is again given by the conditional expectation of the terminal price under the induced dynamics
\[
r^*_{\mathrm{PO}}(t,s)
= \mathbb{E}_{t,s}[S_T]
= s + \epsilon \mu \frac{(T-t)^2}{2}.
\]
Substituting this expression into the standard optimal quoting formulas yields the stated expressions for the bid and ask premia.
\end{proof}

\paragraph{Economic interpretation.}
The performatively optimal strategy internalises the fact that the MM’s own quoting behaviour induces a drift in prices.
Depending on $\epsilon$ and the time horizon, the performatively optimal MM may optimally \emph{sell} in situations where the classical optimal strategy \emph{buys}.
This reflects the exploitation of the endogenous price pressure generated by performativity-unaware agents.

\begin{figure}[t] 
  \centering
  \includegraphics[width=\textwidth]{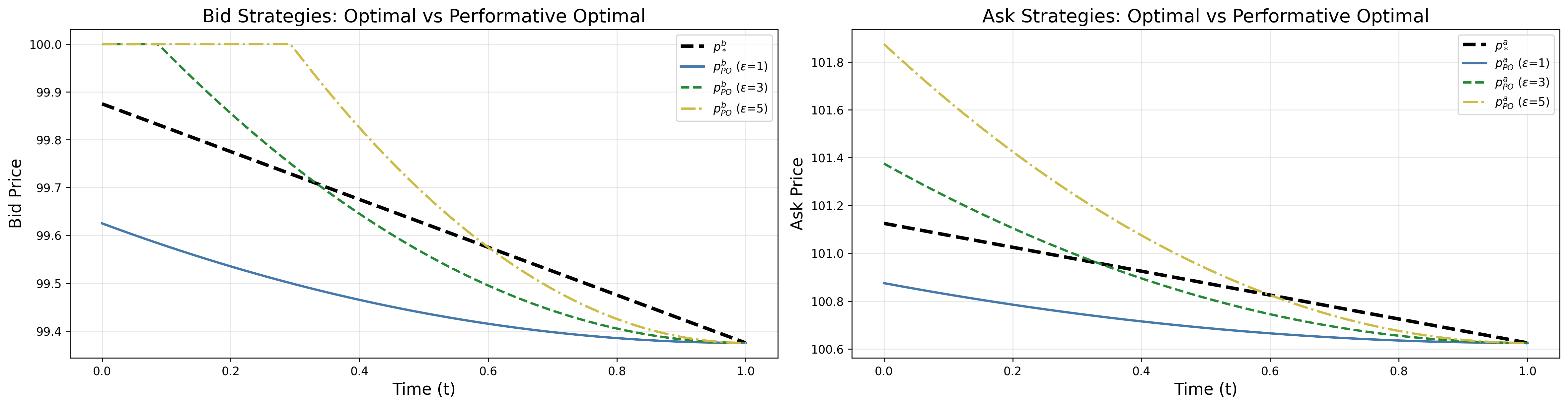}
  \caption{Comparison of optimal bid and ask quotes under classical optimality ($p^b_*,p^a_*$) and performative optimality ($p^b_{\mathrm{PO}},p^a_{\mathrm{PO}}$) for different values of the performativity parameter $\epsilon$.
  In the left panel, $p^b_{\mathrm{PO}}>p^b_*$ indicates a stronger incentive to buy under performative optimality; in the right panel, $p^a_{\mathrm{PO}}>p^a_*$ indicates a stronger incentive to sell.
  The intersection point $p^{a,b}_{\mathrm{PO}}=p^{a,b}_*$ marks a phase transition between buying- and selling-dominant behaviour.
  Parameters: $s=100$, $k=1.6$, $\mu=0.5$, $T=1$.}
  \label{belief_deltas}
\end{figure}

Figure~\ref{belief_deltas} illustrates how performative optimality modifies quoting behaviour relative to the classical optimal strategy.
The figure compares the bid and ask quotes implied by the classical reservation price $r^*$ and the performatively optimal reservation price $r^*_{\mathrm{PO}}$ for different values of the performativity parameter $\epsilon$.

When $\epsilon$ is small, the performative-aware strategy exhibits a stronger tendency to sell, even though the classical strategy $r^*$ is biased toward buying due to the positive drift $\mu$.
The reason is that a performativity-aware market maker correctly anticipates that the price impact induced by her own strategy is weak.
As a result, she exploits the upward pressure generated by performativity-unaware strategies by selling against them.
In this regime, the performative-aware agent effectively arbitrages the naïve belief embedded in $r^*$, which assumes an exogenous drift $\mu$ rather than the true endogenous drift $\epsilon\mu(T-u)$.

As $\epsilon$ increases, the performative effect becomes stronger and the performative-aware strategy adjusts accordingly.
For sufficiently large $\epsilon$, the market maker initially quotes more aggressively on the bid side than the classical strategy, accumulating inventory in anticipation of stronger endogenous price appreciation.
At a critical time, marked in Figure~\ref{belief_deltas} by the intersection $p^{a,b}_{\mathrm{PO}}=p^{a,b}_*$, a phase transition occurs.
Beyond this point, the performative-aware strategy shifts toward selling, placing ask quotes closer to the mid-price in order to unwind inventory at elevated prices.
Larger values of $\epsilon$ delay this transition, reflecting stronger and more persistent performative price pressure.

This comparison highlights a central implication of performative optimality: accounting for endogenous price feedback can reverse or significantly reshape optimal trading incentives relative to classical formulations.
It also motivates the need for a stability concept, since systematic deployment of $r^*_{\mathrm{PO}}$ will itself alter the price dynamics and trigger further rounds of re-optimisation.

\subsection{Performative Stability under Endogenous Beliefs}

Once the performatively optimal strategy $r^*_{\mathrm{PO}}$ is itself deployed systematically, it induces a new price process of the form
\[
ds_u = \epsilon\bigl(r^*_{\mathrm{PO}}(u,s_u) - s_u\bigr)\,du + \sigma\,dW_u,
\]
which in turn alters the optimisation problem faced by the MM.
Repeated re-optimisation therefore generates a sequence of strategies $\{r^{(n)}\}_{n\ge1}$, where
\[
r^{(n+1)} = \OptMM(r^{(n)}).
\]

\begin{observation}
Repeated application of the optimisation operator $\OptMM(\cdot)$ does not, in general, produce a convergent strategy. Unless a fixed point is reached, each re-optimisation step changes the induced price dynamics, so the environment faced by the market maker differs from the one assumed in the previous optimisation.
\end{observation}

This motivates the use of the generally stronger notion of performative stability.

\begin{theorem}[Performatively Stable Strategy under Endogenous Beliefs]
\label{thm:ps_beliefs}
Assume linear utility and price dynamics of the form \eqref{eq:price}, and suppose that the conditional expectation $\mathbb{E}_{t,s}[S_T]$ is affine in $s$. Then, the market-making optimisation problem admits a suboptimal solution for its value function $H$ with a reservation price $r^*_{\mathrm{PS}}$ which is \emph{performatively stable} if and only if it satisfies
\[
r^*_{\mathrm{PS}}(t,s) = \mathbb{E}_{t,s}[S_T]
\]
under the performative price dynamics $ds_u = \epsilon\bigl(r^*_{\mathrm{PS}}(u,s_u) - s_u\bigr)\,du + \sigma\,dW_u$, $u\in[t,T]$. Equivalently, $r^*_{\mathrm{PS}}$ is a fixed point of the market-maker optimisation operator, i.e., $r^*_{\mathrm{PS}} = \OptMM\! \left(r^*_{\mathrm{PS}}\right)$.

The analytical expression of the performatively stable reservation price is given by
\begin{equation*}
r^*_{PS}(t,s)= s
\end{equation*}
\end{theorem}

\begin{proof}[Proof sketch]
Under linear utility, the optimal reservation price coincides with the conditional expectation of the terminal price under the perceived price dynamics.
Performative stability therefore requires the belief used to form this expectation to be consistent with the price dynamics induced by the reservation price itself. Suppose that prices evolve according to
\[
ds_u = \epsilon\bigl(r^*_{\mathrm{PS}}(u,s_u) - s_u\bigr)\,du + \sigma\,dW_u,
\qquad u\in[t,T].
\]
Thus, it must hold $r^{*}_{PS} = \mathbb{E}_{t,s}[S_T]$ with $\mathbb{E}_{t,s}[S_T]$ affine in $s$, which implies that $r^*_{PS}$ admits an affine representation $
r^*_{PS}(t,s)= \alpha(t)+\beta(t)s$ for all $(t,s).$ Substituting this form in the price process we obtain
\begin{align}\label{perf_stable_step}
\mathbb{E}_{t,s}[S_T] =\Phi(t)s + \int_t^T \epsilon\Phi(u)\alpha(u)du, \quad \Phi(u):=\exp\!\left(\int_u^T c(\kappa)\,d\kappa\right),
\quad
c(\kappa):=\epsilon[\beta(\kappa)-1].
\end{align}
Given that $r^*_{PS}$ has the solution $r^*_{PS}=\mathbb{E}_{t,s}[S_T]$, it must be true that:
\[
\alpha(t)+\beta(t)s = \Phi(t)s + \int_t^T \epsilon\Phi(u)\alpha(u)du.
\]
Since this must hold for all s, by matching the coefficients we obtain:
\begin{align}
\notag\beta(t) = 1,\hspace{0.05cm} \alpha(t) = 0  \hspace{0.2cm}\forall t.
\end{align}

The solution $(\alpha(t), \beta(t))$ of Equation \eqref{perf_stable_step} defines a reservation price that is optimal under the price dynamics it induces. Therefore, $r^*_{PS}(t,s)=s$ for all $(t,s)$.


Conversely, any performatively stable reservation price must satisfy this fixed-point condition.
Hence, $r^*_{\mathrm{PS}}$ is a fixed point of the optimisation operator:
\[
r^*_{\mathrm{PS}} = \OptMM(r^*_{\mathrm{PS}}).
\]

Existence and uniqueness of solutions to the integral equation follow from standard arguments under linearity and boundedness; full details are provided in Appendix~\ref{stable_proof_1}. Note that since the operator $\OptMM(\cdot)$ is defined via the solution of the market-making HJB under the induced price dynamics, this fixed-point condition is equivalent to $r^*_{\mathrm{PS}}$ being invariant under repeated HJB optimisation.
\end{proof}

\paragraph{Economic Interpretation.}
Performative stability eliminates arbitrage between performativity-unaware and performativity-aware agents.
It identifies strategies whose induced price dynamics, beliefs, and optimal responses are mutually consistent.
In contrast, repeated re-optimisation without stability leads to endogenous instability and belief-driven decay of profits.

\begin{enumerate}[left=0pt, nosep]
    \item Repeated re-optimisation embeds the time-adjusted belief term into the price dynamics, which performativity-aware agents can systematically exploit. This induces an endogenous strategic arms race in which agents continually adjust their strategies in response to the performative effects of others.

    \item It is interesting and, perhaps surprising, that in this setting, the performatively stable strategy coincides with the martingale benchmark, i.e., the case with zero drift belief. This is because any directional belief becomes detectable through its performative impact and can be arbitraged by performativity-aware agents. \label{martingale}
\end{enumerate}

This simple example illustrates how the performativity framework can be instantiated in closed form and sets the scene for the richer models studied in the subsequent section.
We will show that point~(\ref{martingale}) is, in fact, an artefact of the present setup: in more general models, the performatively stable solution need not coincide with the (trivial) martingale solution.

\section{Performative Market Making under Endogenous Constraints} \label{constraints}
 
The primary source of risk for a market maker arises from the accumulation of inventory through filled orders. Consequently, market-making strategies must incorporate an algorithmic mechanism for inventory control. As we show in the following sections, even in the absence of directional beliefs about the price, the inventory management rule itself can induce a performative effect on the price process, thereby exposing the market maker to risks generated by her own endogenous constraints.

\subsection{Baseline: Market Making under Inventory Constraints}

We consider the classical market-making model of \citet{avellaneda2008} (A\&S), which incorporates inventory control and provides a closed-form optimal strategy. Suppose that the market maker has an exponential utility $\phi(x,s,q)=-\exp{\{-\gamma(x+qs)\}}$ and optimises her strategy under a standard Brownian montion
\begin{align} \label{eq:brown}
     ds_u = \sigma dW_u, \qquad u \in [t,T],
\end{align}
i.e. she holds no directional beliefs about the terminal mid-price, $\mathbb{E}_{t,s}[S_T]=s_t$.

\begin{theorem}[Optimal Reservation Price under Inventory Constraints \cite{avellaneda2008}]
\label{lem:baseline_opt_inv}
Under the price dynamics~\eqref{eq:brown} and exponential utility, the optimal reservation price satisfies
\[
r^*(t,s) = s - \gamma q \sigma^2 (T-t)
\]
where $\gamma>0$ the risk aversion parameter of the MM, $q\in \mathbb{Z}$ the inventory of the asset she holds, and $\sigma>0$ the volatility of the asset. The optimal bid and ask premia are given by
\[
\delta_*^{a,b} = \frac{1}{\gamma} \ln \left( 1 + \frac{\gamma}{k} \right)+\frac{1}{2}\gamma\sigma^2(T-t),
\qquad
\psi^* = \delta_*^a + \delta_*^b = \frac{2}{\gamma} \ln \left( 1 + \frac{\gamma}{k} \right) + \gamma\sigma^2(T-t).
\]
where $k>0$ is the order-book depth parameter that controls the sensitivity of order arrival rates to quoted prices, cf. Equation~ \eqref{eq:intensities}.
\end{theorem}

\noindent 
The optimal quotes are symmetric around the reservation price $r^*(t,s)$, but generally skewed relative to the mid-price $s$ due to inventory. When the inventory is positive, i.e. $q>0$, the reservation price satisfies $r^*<s$, leading to more competitive ask quotes and a tendency to sell; conversely, when $q<0$, the reservation price satisfies $r^*>s$, tightening the bid side and encouraging purchases.


\subsection{Performative Optimality under Endogenous Constraints}

We now assume that the prevailing strategy in the market follows the A\&S model described in Theorem~\ref{lem:baseline_opt_inv}. The resulting price dynamics then follows
\begin{align} \label{perf_as_price}
ds_u 
= \epsilon\left(r^*_u - s_u \right) du + \sigma\, dW_u
= -\epsilon\gamma \sigma^2 q_u (T - u)du + \sigma\, dW_u,
\qquad u \in [t, T],
\end{align}
where $\epsilon>0$ measures the strength of the market’s reaction to the A\&S reservation price. As we see from equation \eqref{perf_as_price}, the mean-reverting term is determined by the inventory-dependent component, $g(u) = -\gamma \sigma^2 q_u (T - u)$, of the A\&S reservation price. Under these dynamics, the conditional expectation of the terminal price satisfies
\begin{align} \label{exp_no_assumption}
\mathbb{E}_{t,s}[S_T] 
= s_t 
-\epsilon \gamma \sigma^2 
\int_t^T (T-u)\mathbb{E}_{t,s}[q_u]\,du.
\end{align}

Since the inventory evolves according to $dq_t = dN^b_t - dN^a_t$ (cf. \Cref{preliminaries})
the expectation in~\eqref{exp_no_assumption} generally does not admit an easy to compute, closed form solution. We therefore introduce the following approximation.

\begin{assumption}[Linear Inventory Approximation]\label{assmpt:inv}
The point processes have symmetric arrival rates intensities $\lambda^a(\delta^a_t) = \lambda^b(\delta^b_t) = A e^{-k \delta}$for $(t,T]$ and therefore it holds $\mathbb{E}_{t,s}[q_u]  =q_t$.\footnote{One could also notice that $(q_t)_t$ is a zero mean-reverting process and model it in that way. However, this would dastrically increase the complexity for $\OptMM$. In fact, this would require to modify the $HJB$ equation which, in turn, would make the analytical solution much more complex than our current approach. As we will see in the simulations, \Cref{assmpt:inv} does not significantly impact the quality of the derived performatively optimal strategy.}
\end{assumption}
 

\noindent 
Under this approximation, Equation~\eqref{exp_no_assumption} becomes
\begin{equation} \label{exp_assumption}
\mathbb{E}_{t,s}[S_T] 
= s_t 
-\epsilon \gamma \sigma^2 q_t\frac{(T-t)^2}{2}.
\end{equation}
A performativity-aware agent with risk parameter $\gamma_{perf}$ and inventory $q_{perf}$ can now optimise their strategy with respect to the (performative) price distribution \eqref{perf_as_price} induced by the prevailing A\&S inventory management.

\begin{proposition}[Performatively Optimal Strategy]
\label{prop:po_inventory} Under Assumption~\ref{assmpt:inv}, a linear approximation of ~\eqref{eq:intensities}\footnote{That is a first order Taylor expansion of the trading intensities: $e^{-k\delta^{a,b}}=1-k\delta^{a,b}+...$. This is a canonical assumption of the $\OptMM$ operator, see., e.g., \cite{avellaneda2008, fodra2012, cartea_book, guent_book}.}, and the performative price process~\eqref{perf_as_price}, a performativity-aware agent with exponential utility, risk aversion $\gamma_{perf}$, and inventory $q_{perf}$
has the following performatively optimal reservation price
\[
r^*_{\mathrm{PO}}(t,s) = \OptMM(r^*)
= s - \sigma^2(T-t)\left(\epsilon \gamma q \frac{(T-t)}{2}+\gamma_{perf}q_{perf}\right)
\]

\noindent The corresponding optimal bid and ask premia satisfy
\begin{align}
&\delta_{PO}^{a,b} = \frac{1}{\gamma_{perf}}\log\!\left(1+\frac{\gamma_{perf}}{k}\right)
+\frac{1}{2}\gamma_{perf}\sigma^2(T-t) \mp \sigma^2(T-t)\left(\epsilon \gamma q \frac{(T-t)}{2}+\gamma_{perf}q_{perf}\right)\\
& \psi^*_{PO} = \delta_*^a + \delta_*^b = \frac{2}{\gamma_{perf}}\log\!\left(1+\frac{\gamma_{perf}}{k}\right) + \gamma_{perf}\sigma^2(T-t).
\end{align}
\end{proposition}

\begin{proof}
Under~\eqref{perf_as_price} and Assumption~\ref{assmpt:inv}, the conditional expectation~\eqref{exp_assumption} is affine in $s$, with
\[
\alpha(t)
=
-\epsilon \gamma \sigma^2 q_t\frac{(T-t)^2}{2},
\qquad
\beta(t)=1,
\]
and $\sigma(t,s)=\sigma(t)$. The assumptions of the operator $\OptMM$ for exponential utility are therefore satisfied (see Theorem~\ref{lem:generic_opt_solutions} in the appendix). Direct substitution (see also appendix) into the solution form of \Cref{lem:generic_opt_solutions}, i.e., performing the optimisation step $r^*_{PO}=\OptMM(r^*)$ for the performativity-aware agent yields the stated expressions for $r^*_{PO},\; \delta_{PO}^{a,b}$ and $\psi^*_{PO}$.
\end{proof}

\paragraph{Interpretation of the Performatively Optimal MM Policy}

The performativity-aware agent’s inventory $q_{perf}$ is affected by her own quoting controls $\delta^{a,b}_{PO}$, whereas she has no control over the prevailing inventory $q$; this enters her optimisation only as an exogenous adjustment term in the reservation price.

\begin{figure}[t] 
  \centering
  \includegraphics[width=\textwidth]{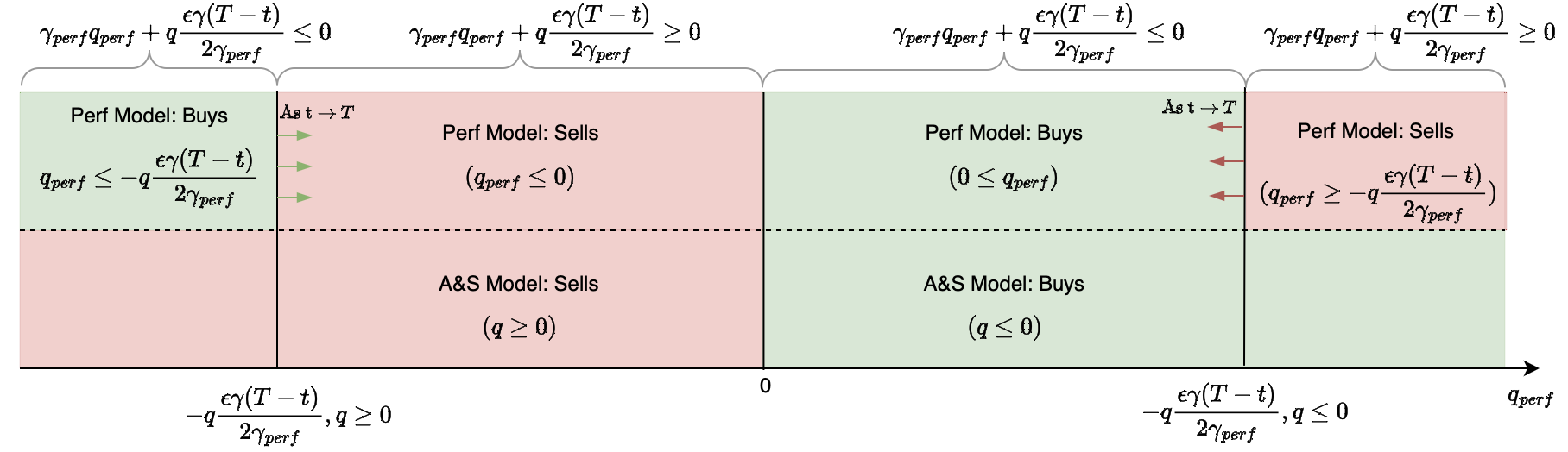}
  \caption{The area above the horizontal dashed line indicates the decisions of the Performative Aware MM strategy and the area below it the decisions of the original A\&S strategy. When the areas below and above have the same colour, the decisions align and when not, the performative MM is arbitraging the prevailing strategy. Her behaviour changes not only based on the sign of her inventory (like the A\&S strategy) but also on its relative value with respect to the critical points shown on the x-axis.}
  \Description{The area above the horizontal dashed line indicates the decisions of the Performative MM and the line below it the decisions of the original A\&S strategy. When the areas below and above have the same colour, the decisions align and when not, the performative MM is arbitraging the prevailing strategy.}
  \label{hand_diagram}
\end{figure}

This implies that the direction of her quoting behaviour is determined by the sign of the resulting reservation-price adjustment. When the reservation price lies below (above) the mid-price, the agent prefers to sell (buy). In the baseline A\&S strategy, this preference is determined solely by the sign of $q$. By contrast, for the performativity-aware agent, this preference, i.e. the sign of the reservation-price adjustment, is governed by the term $\left[\epsilon \gamma q \frac{(T-t)}{2}+\gamma_{perf}q_{perf}\right]$. Thus, the optimal quoting decision depends on both the prevailing inventory $q$ and the agent’s own inventory $q_{perf}$. This induces critical thresholds for $q_{perf}$ at which the performativity-aware agent undergoes strategic phase transition in her quoting behavior, governed by the time decaying term $-|q|\frac{\epsilon \gamma(T-t)}{2\gamma_{perf}}$, as illustrated in Figure \ref{hand_diagram}. The resulting thresholds define regions in which the performativity-aware agent either aligns with or trades against the prevailing A\&S strategy.

In particular, when the prevailing inventory induces predictable price drift, the performativity-aware agent may deliberately accumulate inventory in anticipation of this drift and subsequently unwind her position as the induced price movement materialises. This mechanism illustrates how predictable price dynamics generated by prevailing inventory control rules can create transient arbitrage opportunities for agents that account for performative effects.\par

\begin{example}
    Consider, for example, the case when $q\leq0$ and $q_{perf}\geq0$. Despite already holding a non-negative inventory, the performative MM continues to accumulate inventory until the point $|q|\frac{\epsilon \gamma(T-t)}{2\gamma_{perf}}$ is reached. At the same time, the A\&S strategy also buys the asset, but because of lack of inventory; $q\leq 0$ means she was short selling before. Beyond the point $|q|\frac{\epsilon \gamma(T-t)}{2\gamma_{perf}}$, the perfomative MM's behaviour becomes even more strategic. She now anticipates the positive deterministic drift in the asset's price induced by the A\&S strategy which signals MMs to buy the asset (when $q\leq0$) and holds sufficient inventory to sell and profit from the increasing demand. Thus, with sufficient inventory, the performative MM continues to sell throughout the time that the price rises, up until other MMs shift their quoting behaviour as their inventories increase and eventually turn positive. In other words, the performative MM arbitrages the price until it stops rising due to the desire of the MMs to buy. This highlights how she exploits predictable price dynamics caused by non-performative (here A\&S) agents. Similar observations can be made for the symmetric case where  $q_{perf} \leq0$ and $q\geq0$.
\end{example}

These results provide a formalisation of qualitative insights on performativity in financial markets, as discussed in \cite{MacKenzie2006}. In particular, the model captures the self-fulfilling feedback between quotes and prices and shows how it can generate temporary arbitrage opportunities for performativity-aware agents. In Section~\ref{Simulations}, we validate these predictions through numerical experiments.

\subsection{Performative Stability under Endogenous Constraints}

In this section, we show 
that a non trivial performative stable strategy exists in this context.  

\begin{theorem}[Performatively Stable Strategy under Endogenous Constraints]
\label{thm:ps_inventory}
Assume exponential utility and price dynamics of the form \eqref{eq:price}, and suppose that the conditional expectation $\mathbb{E}_{t,s}[S_T]$ is affine in $s$. Then, under Assumption \ref{assmpt:inv}, the market-making optimisation problem admits a suboptimal solution for its value function $H$ with a reservation price $r^*_{\mathrm{PS}}$ which is \emph{performatively stable} if and only if it satisfies
\[
r^*_{\mathrm{PS}}(t,s) = s-q\gamma\sigma^2\frac{(e^{\epsilon(T-t)}-1)}{\epsilon}
\]
under the performative price dynamics $ds_u = \epsilon\bigl(r^*_{\mathrm{PS}}(u,s_u) - s_u\bigr)\,du + \sigma\,dW_u$, $u\in[t,T]$. 
\end{theorem}

\begin{proof}[Proof sketch]
For the derivation of the analytical expression of $r^*_{PS}(t,s)$, arguments similar to the proof of \Cref{thm:ps_beliefs} hold. 
For the full derivation, we refer the reader to Appendix \ref{proof:ps_as}. We here present the verification that the strategy is indeed stable. 

Consider the strategy $r^*(t,s):=s-q\gamma\sigma^2\frac{(e^{\epsilon(T-t)}-1)}{\epsilon}$. Then the midprice evolution under the performative impact of this strategy is given by the equation:
\[
ds_u = \epsilon(r^*(u,s)-s)du + \sigma dW_u = - q\gamma\sigma^2(e^{\epsilon(T-t)}-1)du + \sigma\,dW_u
\]
which yields:
\[
\mathbb{E}_{t,s}[S_T] =s_t - q_t\gamma\sigma^2\left(\frac{e^{\epsilon(T-t)}-1}{\epsilon}-(T-t)\right).
\]
Notice that the midprice dynamics is of the form of \eqref{eq:price} and that the conditional expectation is affine to s. Therefore assuming exponential utility, the linear approximation of the trading intensities and Assumption 1, we can use the $\OptMM(r^*)$ operator to derive the performatively optimal reservation price $r^*_{PO}$. Direct substitution of our setup's configuration into the solution form of \Cref{lem:generic_opt_solutions} yields that $r^*_{PO}=\OptMM(r^*)=s-q\gamma\sigma^2\frac{(e^{\epsilon(T-t)}-1)}{\epsilon}=r^*$. Therefore $r^*=r^*_{PS}(t,s)$.

Existence and uniqueness of the solution follow standard arguments under linearity and boundedness; full details are provided in Appendix~\ref{proof:ps_as}. Note that since the operator $\OptMM(\cdot)$ is defined via the solution of the market-making HJB under the induced price dynamics, this fixed-point condition is equivalent to $r^*_{\mathrm{PS}}$ being invariant under repeated HJB optimisation.
\end{proof}

\paragraph{Economic Interpretation.} The derivation of the performatively stable reservation price $r_{PS}^*$ provides several critical insights into the long-run behaviour of markets where strategies and prices are co-dependent.

\begin{enumerate}[left=0pt, nosep]
    \item \emph{Incentive Compatibility and Equilibrium:} Performative stability identifies a state of consistency where the market maker's belief about price evolution is perfectly aligned with the price dynamics induced by their own strategy. In this state, the agent has no incentive to further re-optimize, as their current model is already the best response to the environment it creates.
    \item \emph{Elimination of Exploitable Drift:} Unlike the performatively optimal strategy, which seeks to exploit the predictable drift created by unaware agents, the stable strategy eliminates these transient arbitrage opportunities. By incorporating the term $\frac{e^{\epsilon(T-t)}-1}{\epsilon}$ into the reservation price, the MM internalizes the impact of their inventory management, resulting in a strategy that is robust to its own performative effects.
    \item \emph{Resolution of the Strategic Arms Race:} Without reaching a stable fixed point, the iterative process of re-optimization can lead to endogenous instability and a decay in profits as agents continually chase distribution shifts. Theorem \ref{thm:ps_inventory} proves that a unique fixed point exists even under complex inventory constraints, providing a terminal solution to this recursive strategic loop.
    \item \emph{Impact of Sensitivity:} The presence of the performativity parameter $\epsilon$ in the stable solution $r_{PS}^*$ indicates that as the market's conformity to the model increases, the MM must adjust their quoting skew more aggressively than in classical non-performative settings to maintain the same level of risk-adjusted optimality.
\end{enumerate}


\section{Simulations} \label{Simulations}

In this section\footnote{For reproducibility, our code repository can be found here: https://github.com/kleitsikas/Performative-Market-Making}, we will focus on the results presented in Section \ref{constraints} 
as their baseline is based on the seminal model of \cite{avellaneda2008}, considered a standard MM model both in academia and industry. The goal of this section is to.
empirically examine and demonstrate the difference between the actual profits (PnL) of the original A\&S model ($r^*_{A\&S}$) and their best response counterparts we provided in Proposition \ref{prop:po_inventory} ($r^*_{PO}$) under the former's performative effects.

Towards that goals, in this section, we first show how market prices evolve under performative effects. Then, we present a comprehensive set of simulations comparing strategies in different scenarios. We evaluate their performance using various metrics across a broad range of $\epsilon$'s and different risk aversion profiles.

\subsection{Price Formation}

For the simulations we consider a discrete time form of the the process $ds_u= \epsilon(r^*_{A\&S}-s_u)du+\sigma dW_u$ which is: 
\begin{equation} \label{eq14}
s_{n+1} = s_n  - \epsilon \gamma \sigma^2 q_n (T - t_n)\Delta t + \sigma \sqrt{\Delta t} \cdot Z_n,
\end{equation}
where $s_n = s(t_n)$, $\Delta t =t_{n+1}-t_{n}$, $Z_n\sim \mathcal{N}(0,1)$ are i.i.d. standard normal random variables. The first two terms represent the deterministic part of our process, and the third term represents the random noise. 

\begin{wrapfigure}{r}{0.5\textwidth}
  \centering
  \includegraphics[width=0.45\textwidth]{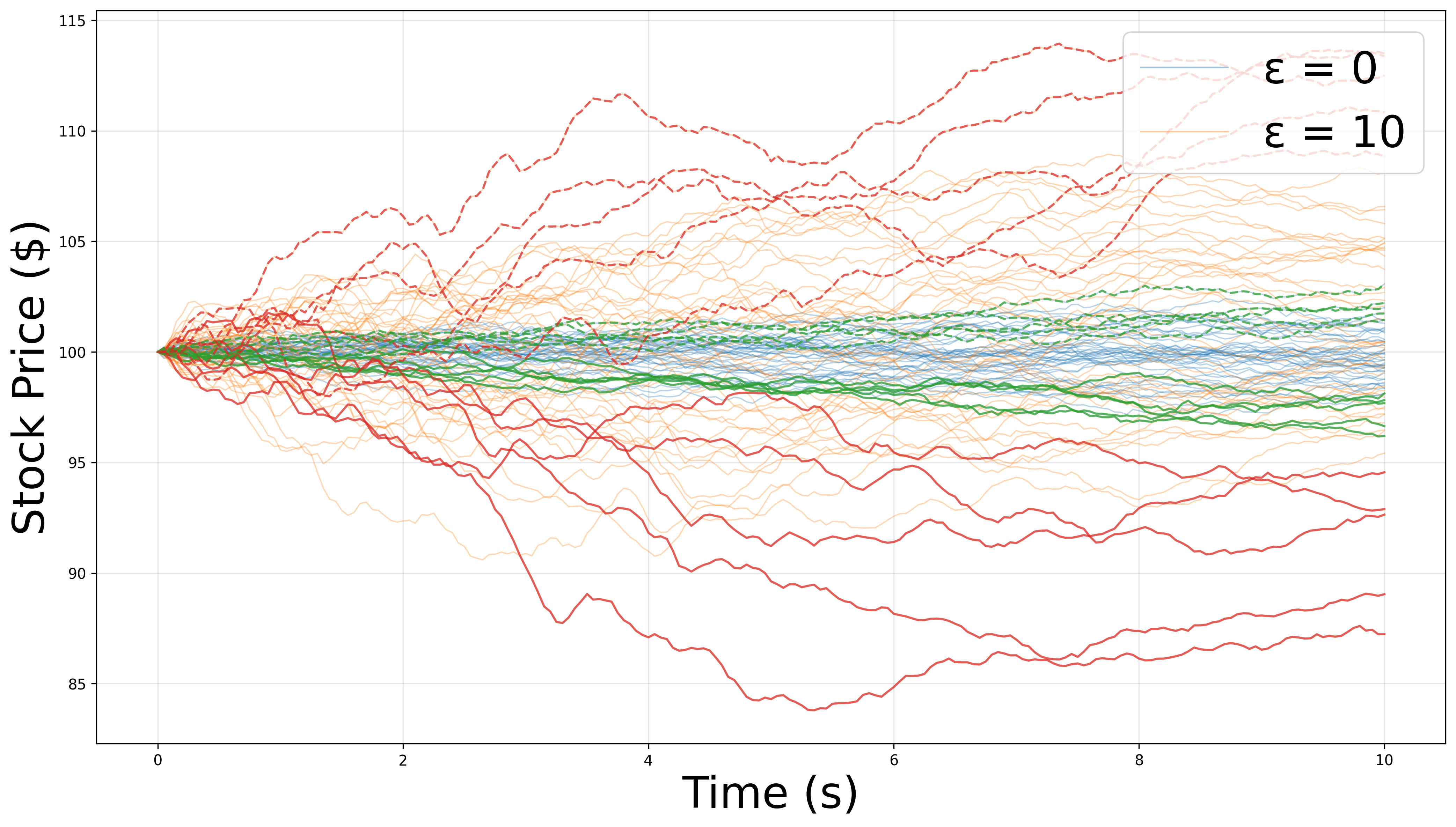}
  \caption{The shaded red price paths are the result of the performative effect that the MM model of A\&S has to the price due to its inventory management ($\epsilon=10$). The blue price paths are the result of pure Brownian noise ($\epsilon=0$). The highlighted dashed and solid lines for each case showcase approximately 10\% of the most price paths, i.e., those with maximum and minimum final prices. The price paths were produced for $s_0=100$, $T=10$, $\gamma=0.5$, $\sigma=0.4$, $\Delta_t=0.05$.}
  \label{price_paths_comparison}
\end{wrapfigure}

To build a better intuition about performativity we plot 50 price paths generated by Equation \eqref{eq14} when $\epsilon=10$ and 50 price paths of pure brownian noise ($\epsilon=0$) in the same graph to demonstrate their fundamental differences, see Figure \ref{price_paths_comparison}. We see how the performative effect of 
the A\&S strategy and its inventory management are the main driver of the price for the shaded red price paths, which capture the conformity towards the A\&S's reservation price. 

We note, interestingly, that these performative fluctuations can be misinterpreted as noise by unaware MMs, especially in  high-frequency trading. However, not only are they not random, but they are fundamentally determined by the prevailing financial model; in this case, the A\&S's inventory adjustment strategy, and ignoring them can result in systematic losses. As we show next, the performative-aware agent can distinguish the model impact from actual noise and capitalize on that to systematically make profits.

\subsection{Simulation Setup}
For the simulations, we set the parameters as $T=10$, $\sigma=0.4$, $A=15$, $k=1.6$, $\Delta t=0.05$, $q_0=0$. For the MMs' risk-aversion profile, $\gamma$, we consider three representative cases: $\gamma=\gamma_{perf}=0.2$ (high-risk), $\gamma=\gamma_{perf}=0.5$ (medium-risk), and $\gamma=\gamma_{perf}=0.8$ (low-risk). The risk profile of the original A\&S strategy influences her inventory pressure and, thus, the price itself. For the performative parameter $\epsilon$, we consider 20 points in the range $[0,20]$ covering weak to  very strong mean-reversion speeds.\par
Each simulation unfolds as follows: at each time step $t$ (equivalently $n$), the strategy computes quotes $\delta^{a}_n$, $\delta^b_n$ and the state variables are updated for $t+\Delta t$ (equivalently $n+1$). The quotes are filled probabilistically based on the intensities from Equation \eqref{eq:intensities}. If $\delta^a_n$ is filled, the agent's inventory, $q_{n+1}$, decreases by one and her wealth in cash changes by $s+\delta^a_n$. Analogously, if $\delta^b_n$ is filled, $q_{n+1}$ increases by one and the agent's wealth in cash changes by $s-\delta^b_n$. We assume that the agents can short stock ($q<0$). The P\&L is updated as $PnL_{n+1} = x_{n+1} + q_{n+1} \cdot s_n$. Theoretically, the solution of the $\OptMM$ operator allows for $\delta^a$, $\delta^b$ $\leq 0$; for our strategies, this means that the agent will execute a market order. These simulation dynamics are used for the comparison of three distinct, which we summarize below.

\begin{description}[leftmargin=*,align=left,nosep, font=\normalfont]
    \item[\underline{\em Inventory strategy ($r^*_{A\&S}$).}] The A\&S model uses the reservation price spread from Theorem \ref{lem:baseline_opt_inv}.

    \item[\underline{\em Symmetric Strategy ($r_{sym}$).}] Quotes are placed symmetrically around the mid-price without an opinion on the asset's evaluation (equivalently $r_{sym}=s$). The strategy's spread uses the term $\frac{2}{\gamma} \ln \left( 1 + \frac{\gamma}{k}\right)+\gamma\sigma^2(T-t)$, that is the same between both the A\&S model and our performative optimal strategy (see Theorems \ref{lem:baseline_opt_inv} and \ref{prop:po_inventory}). This is a natural choice for a fair comparison.
    \item[\underline{\em Performativity-aware Strategy ($r^*_{PO}$).}] This 
    strategy 
    uses the equations of Theorem \ref{prop:po_inventory}.
\end{description}

\noindent A strategy's performance is primarily assessed by its P\&L profile, and inventory management. Strategies that achieve higher terminal wealth with lower variance and reduced inventory exposure are typically considered desirable. Accordingly, we consider the following evaluation metrics.

\begin{description}[leftmargin=*,align=left,nosep,font=\normalfont]
    \item[\underline{\em Profit and Loss (P\&L), Sharpe Ratio.}] The P\&L distribution profile of a trading strategy is the main metric to evaluate its performance. We focus on the distribution of the agent’s terminal wealth at the end of the simulations. We record the mean and variance of the P\&L profiles of the simulations as well as their Sharpe Ratio.
    \item[\underline{\em Terminal Inventory.}] This metric captures the outstanding position held by the MM at the end of the trading session. Ideally, a market maker aims to conclude the trading session with a terminal inventory close to zero, though this is not always a strict constraint.
\end{description}

To compute the above metrics for the strategies, we run $1,000$ simulations for each parameter combination.

\subsection{Results}

\begin{figure}[t]
  \centering\vspace{-0.2cm}
  \includegraphics[width=0.33\linewidth, trim=10 0 40 0, clip]{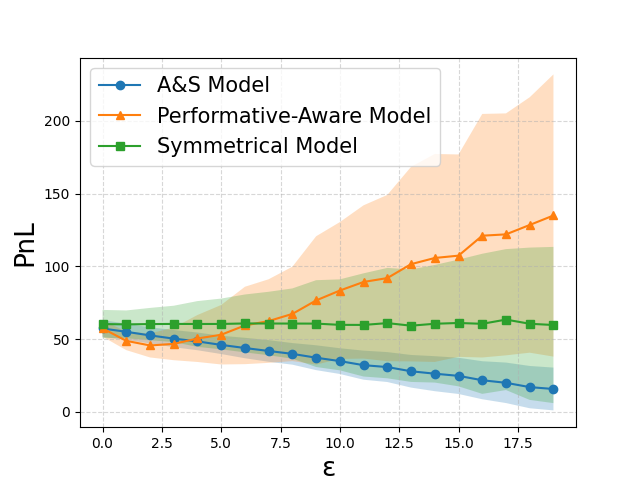}%
  \hfill
  \includegraphics[width=0.33\linewidth, trim=10 0 40 0, clip]{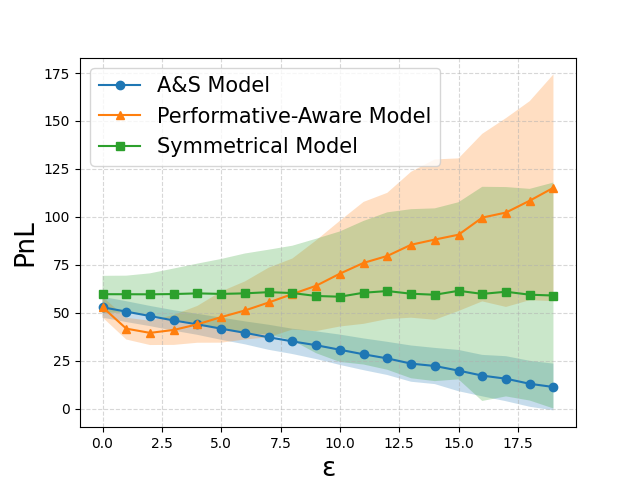}%
  \hfill
  \includegraphics[width=0.33\linewidth, trim=10 0 40 0, clip]{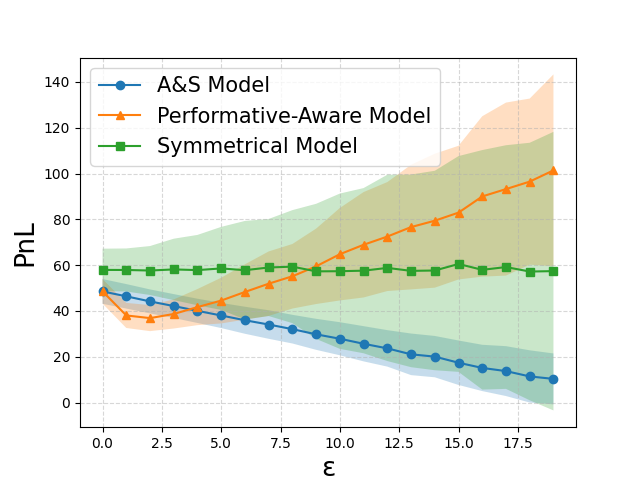}
  \caption{The PnL of all strategies across all $\epsilon$ values for $\gamma=0.2, 0.5$ and $0.8$. Every point is the average of 1000 runs and the shaded area is $\pm$ 1 standard deviation from the mean. Performative strategies consistently outperform the A\&S strategy with higher PnL across all $\gamma$'s and the symmetric strategy after $\epsilon$ surpasses a break-off point.}\vspace{-0.3cm}
  \label{fig:scenario1_results}
  \Description{PnL of all strategies.}
\end{figure}

We are examining the original A\&S model ($r^*_{A\&S}$) and their best response counterpart we provided in Proposition \ref{prop:po_inventory} ($r^*_{PO}$) under the former's performative effects alongside the symmetric strategy as an additional benchmark. Our main results are illustrated in Figure \ref{fig:scenario1_results}. We assuming for simplicity that all strategies have the same risk profiles $\gamma$\footnote{For more striking differences between the performances, one could fine tune their distinct risk parameter $\gamma_{perf}$.}. We notice a similar behaviour; in all cases, the performative strategy exhibits superior P\&L profiles compared to the benchmarks. With respect to our A\&S baseline, we notice that this happens for values bigger than $\epsilon=4$. Beyond that point, the signal to noise ratio of the performative impact of the A\&S model is strong enough to be exploited by performative aware agents. Importantly, we see that across $\epsilon$'s, the larger the performativity in the market the more misspecified the underlying model (in this case A\&S) that induces it becomes. The $r^*_{PO}$ strategy surpasses the symmetric strategy's PnL after a certain break-off point. Naturally, the lower $\gamma$ the bigger the possible maximum P\&L that the performative  agent can have, as A\&S trades more. Lastly, the fact that the P\&L of $r^*_{PO}$ increases (almost) linearly with $\epsilon$ in contrast to the deteriorating performance 
of A\&S and the constant performance of a symmetric 
shows that the model does not just passively accumulate inventory but acts strategically just like we theoretically concluded in Section \ref{constraints}. \par

To further support our findings, we present the P\&L, Sharpe Ratio and terminal inventory for the case of $\gamma=0.5$ and representative values of $\epsilon$ which can be found in the Appendix (see Table \ref{tab:strategy_results}). 


\section{Conclusions and Future Work} \label{conclusions}
In this paper, we introduced a formal mathematical framework to model performativity in financial markets. By revisiting the traditional use of diffusion processes, we established that financial models do not merely analyze markets but actively shape them through a closed feedback loop. We instantiated this framework within the market-making problem, introducing to this realm the solution concepts of performative optimality and performative stability. 

Our results demonstrate that a performatively stable strategy exists as a fixed point where the model remains invariant under re-optimization. This state eliminates the strategic ``arms race'' that occurs when agents continually adjust to the distributional shifts caused by their own models. Furthermore, our simulations confirm that agents who account for these endogenous feedback loops can significantly outperform performativity-unaware benchmarks by exploiting predictable price dynamics.

This work opens several directions for research at the intersection of Economics and Computation.
A first natural extension is performative mechanism design, i.e., investigate how market-matching rules can be designed to minimize the manipulative potential of performativity-aware agents or to induce socially desirable stable states.
Secondly, while we establish the existence of stability for a dominant agent, investigating the convergence properties of multiple competing, aware agents remains an open question for algorithmic game theory.
Thirdly, as AI-driven models increasingly shape liquidity, it is crucial to understand how price dynamics change when $r(t)$ depends on the parameters of an ML model or the policy of an RL algorithm.
Finally, future research could instantiate this framework for other financial instruments, such as option pricing, to study the systemic effects of performativity across different asset classes.




%
%
%
%
%

\bibliographystyle{ACM-Reference-Format}
\bibliography{bibliography}

\appendix
\section{Extended Proofs for Theoretical Results}

\subsection{Proof of Performative Stable Strategy under Endogenous Beliefs} \label{stable_proof_1}
Under linear utility, the optimal reservation price coincides with the conditional expectation of the terminal price under the perceived price dynamics.
Performative stability therefore requires that the belief used to form this expectation be consistent with the price dynamics induced by the reservation price itself. Suppose that prices evolve according to
\[
ds_u = \epsilon\bigl(r^*_{\mathrm{PS}}(u,s_u) - s_u\bigr)\,du + \sigma\,dW_u,
\qquad u\in[t,T].
\]
Under the assumptions of the $Opt_{MM}$ operator it must hold $r^*_{PS} = \mathbb{E}_{t,s}[S_T] = \alpha(t)+\beta(t)s \forall t,s$, i.e. $\mathbb{E}_{t,s}[S_T]$ is affine to $s$. Substituting this form to the above SDE we obtain:
\begin{align*}
ds_u &= \epsilon\left[\alpha(u)+\beta(u)s_u - s_u\right]\,du + \sigma\,dW_u \\
&=(\epsilon \alpha(u)+\epsilon\left[\beta(u)-1)s_u\right]du+ \sigma\,dW_u\xRightarrow{c(u):=\epsilon(\beta(u)-1)}\\
&= \left[\epsilon \alpha(u)+c(u)s_u\right]du+ \sigma\,dW_u
\end{align*}
which after rearranging terms yields the equation
\begin{align*}
ds_u - c(u)s_udu=\epsilon\alpha(u)du+ \sigma\,dW_u\,.
\end{align*}
By setting $\Phi(u):=\exp{\left(\int_u^T c(\kappa)\,d\kappa\right)}$, we have that $\frac{d\Phi(u)}{du}=-c(u)\Phi(u)$ which gives
\begin{align*}
\Phi(u)ds_u+d\Phi(u)s_u&=\epsilon\Phi(u)\alpha(u)du+ \Phi(u)\sigma\,dW_u\\
d(\Phi(u)s_u) &= \epsilon\Phi(u)\alpha(u)du+ \Phi(u)\sigma\,dW_u \xRightarrow{\Phi(T)=1}\\
S_T &= \Phi(t)s_t + \int_t^T \epsilon\Phi(u)\alpha(u)du + \int_t^T \Phi(u)\sigma\,dW_u\\
\mathbb{E}_{t,s}[S_T] &= \Phi(t)s_t + \int_t^T \epsilon\Phi(u)\alpha(u)du 
\end{align*}
Thus, for the performative stable strategy it must be true that
\begin{align} 
\notag \alpha(t)+\beta(t)s_t = \Phi(t)s_t + \int_t^T \epsilon\Phi(u)\alpha(u)du
\end{align}
where $\Phi(\cdot)$ is the propagator of the SDE.
Since this must hold for all s, by matching the coefficients we obtain that
\begin{align*}
\beta(t)s_t &= \Phi(t)s_t\Rightarrow \beta(t)= \exp{\left(\int_t^T c(\kappa)\,d\kappa\right)}\Rightarrow\\
\beta(t)&=\exp{\left(\int_t^T \epsilon[\beta(\kappa)-1]\,d\kappa\right)}\xRightarrow{\log{}}\\
\log{\beta(t)}&=\int_t^T \epsilon[\beta(\kappa)-1]\,d\kappa\xRightarrow{\frac{\partial}{\partial t}}\\
\frac{\beta'(t)}{\beta(t)} &= \epsilon[1-\beta(t)], \quad\beta(T)=1
\end{align*}
which is a logistic Differential Equation and given that $\beta(T)=1$ is an equilibrium solution for the equation, it holds by uniqueness that $\beta(t)=1$, for all $t>0$. And
\begin{align}
\notag&\alpha(t)=\int_t^T \epsilon\Phi(u)\alpha(u)du, \alpha(T)=0\xRightarrow{\frac{\partial}{\partial t}}\alpha '(t)=-\epsilon \Phi(t)\alpha(t)=-\epsilon\alpha(t)
\end{align}
which with $\alpha(T)=0$ gives $\alpha(t)=0$ for all $t>0$. The solution $(\alpha(t), \beta(t))$ of equation \eqref{perf_stable_step} defines a reservation price that is optimal under the price dynamics it induces.\par

Conversely, any performatively stable reservation price must satisfy this fixed-point condition. Hence, $r^*_{\mathrm{PS}}$ is a fixed point of the optimisation operator:
\[
r^*_{\mathrm{PS}} = \OptMM(r^*_{\mathrm{PS}}).
\]

\subsection{Operator's $\OptMM$ Solution Set under Exponential Utility}

\begin{theorem}[Extended Solution Set of Optimal Reservation Price under Inventory Constraints \cite{fodra2012}] \label{lem:generic_opt_solutions} 
Assume exponential utility, the price dynamics of the form \eqref{eq:price},with its conditional expectation $\mathbb{E}_{t,s}[S_T]=\alpha(t)+\beta(t)s$, i.e. affine in s, $\sigma(t,s(t))=\sigma(t)$, and a linear approximation of the trading intensities in \eqref{eq:intensities}\footnote{i.e. using a first order Taylor expansion of the trading intensities \cite{avellaneda2008},\cite{fodra2012}}. Then the optimal reservation price satisfies
\[
r^*(t,s) = \theta_1 + 2q\theta_2.
\]
The optimal bid and ask premia are given by
\[
\delta_*^{a,b} = \frac{1}{\gamma}\log\!\left(1+\frac{\gamma}{k}\right)
- \theta_2 \pm \bigl(\theta_1 - s + 2q\theta_2),
\qquad \psi^* = \delta_*^a + \delta_*^b = \frac{2}{\gamma}\log\!\left(1+\frac{\gamma}{k}\right) - 2\theta_2
\]
\textit{where}
\begin{align*}
\theta_1(t,s) = \mathbb{E}_{t,s}\!\bigl[S(T)\bigr],\qquad
\theta_2(t)
= - \frac{1}{2}\gamma \int_t^T \sigma^2(\xi)\beta^2(\xi,T)\,d\xi.
\end{align*}
\end{theorem}

\subsection{Performing the $r^*_{PO}=\OptMM(r^*)$ Step (Proof of Proposition \ref{prop:po_inventory})}

Under Assumption \ref{assmpt:inv}, a linear approximation of trading intensities in equation~\eqref{eq:intensities}, and the performative price process~\eqref{perf_as_price}, a performativity-aware agent with exponential utility, risk aversion $\gamma_{perf}$, and inventory $q_{perf}$ satisfies the conditions for the formulas of \Cref{lem:generic_opt_solutions} for which we notice that 

\begin{align*}
\theta_1(t,s) = s -\epsilon \gamma \sigma^2 q\frac{(T-t)}{2},\qquad
\theta_2(t)
= - \frac{1}{2}\gamma_{perf}\sigma^2(T-t)
\end{align*}

Therefore the performatively optimal reservation price becomes:
\begin{align*}
r^*_{\mathrm{PO}}(t,s)
&= \underbrace{s -\epsilon \gamma \sigma^2 q\frac{(T-t)^2}{2}}_{\theta_1(t,s)} + 2q_{perf}\left(\underbrace{- \frac{1}{2}\gamma_{perf}\sigma^2(T-t)}_{\theta_2(t)}\right)
\\&= s - \sigma^2(T-t)\left(\epsilon \gamma q \frac{(T-t)}{2}+\gamma_{perf}q_{perf}\right).
\end{align*}

\noindent The corresponding optimal bid and ask premia satisfy
\begin{align}
\notag\delta_{PO}^{a,b} &= {\gamma_{perf}}\log\!\left(1+\frac{\gamma_{perf}}{k}\right) - \theta_2 \pm \bigl(\theta_1 - s + 2q\theta_2)\\\notag &= \frac{1}{\gamma_{perf}}\log\!\left(1+\frac{\gamma_{perf}}{k}\right)
+\frac{1}{2}\gamma_{perf}\sigma^2(T-t) \pm (r^*_{PO}-s)=\\
\notag&=\frac{1}{\gamma_{perf}}\log\!\left(1+\frac{\gamma_{perf}}{k}\right)
+\frac{1}{2}\gamma_{perf}\sigma^2(T-t)\pm \left(- \sigma^2(T-t)\left(\epsilon \gamma q \frac{(T-t)}{2}+\gamma_{perf}q_{perf}\right)\right)\\
\notag \psi^*_{PO} &= \delta_*^a + \delta_*^b = \frac{2}{\gamma_{perf}}\log\!\left(1+\frac{\gamma_{perf}}{k}\right) + \gamma_{perf}\sigma^2(T-t).
\end{align}

\subsection{Proof of Performative Stable Strategy under Endogenous Constraints} \label{proof:ps_as}

Assume exponential utility and price dynamics of the form \eqref{eq:price}, and suppose that the conditional expectation $\mathbb{E}_{t,s}[S_T]$ is affine in $s$. Then, under Assumption \ref{assmpt:inv}, the market-making optimisation problem admits a suboptimal solution for its value function $H$ with a reservation price $r^*_{\mathrm{PS}}$ which holds Theorem's \ref{lem:generic_opt_solutions} solution set formula:

\[
r^*_{\mathrm{PS}}(t,s) = \theta_1(t,s)+2q\theta_2(t)
\]
\begin{align*}
\theta_1(t,s) = \mathbb{E}_{t,s}\!\bigl[S(T)\bigr],\qquad
\theta_2(t)
= - \frac{1}{2}\gamma\sigma^2 \int_t^T \beta^2(\xi,T)\,d\xi.
\end{align*}

Given that $\mathbb{E}_{t,s}[S_T]= \alpha(t)+\beta(t)s$,
$\theta_1(t,s)=\alpha(t)+\beta(t)s$.
Therefore \[r^*_{PS}=\alpha(t)+\beta(t)s+2q\theta_2(t)=a(t)+2q\theta_2(t)+\beta(t)s.
\]
Substituting this to the following equation
\[
ds_u = \epsilon\bigl(r^*_{\mathrm{PS}}(u,s_u) - s_u\bigr)\,du + \sigma\,dW_u,
\qquad u\in[t,T].
\]
we get:
\[
ds_u = \epsilon\bigl(a(t)+2q\theta_2(t)+\beta(t)s - s_u\bigr)\,du + \sigma\,dW_u
\]
\[
ds_u = \bigl(\epsilon(a(u)+2q_u\theta_2(u))+\epsilon(\beta(u)-1)s_u \bigr)\,du + \sigma\,dW_u
\]
We notice that 
\[\epsilon(\beta(t)-1)=c(u)\] as in \ref{stable_proof_1}
Also if we define $A(u):=(a(u)+2q_u\theta_2(u))$ the above SDE becomes:
\[
ds_u = \bigl(\epsilon A(u)+c(u)s_u \bigr)\,du + \sigma\,dW_u
\]
Therefore from the Proof of \ref{stable_proof_1}

\begin{align}
S_T &= \Phi(t)s_t + \int_t^T \epsilon\Phi(u)A(u)du + \int_t^T \Phi(u)\sigma\,dW_u\\
\mathbb{E}_{t,s}[S_T] &= \Phi(t)s_t + \int_t^T \epsilon\Phi(u)A(u)du 
\end{align}
where
\begin{align}\label{perf_stable_step}
A(u):=a(u)+2q_u\theta_2(u), \quad \Phi(u):=\exp\!\left(\int_u^T c(\kappa)\,d\kappa\right),
\quad
c(\kappa):=\epsilon[\beta(\kappa)-1].
\end{align}
Given that $r^*_{PS}$ has the solution $r^*_{PS}=a(t)+2q\theta_2(t)+\beta(t)s$, with  $\mathbb{E}_{t,s}[S_T]= \alpha(t)+\beta(t)s$
it must be true that:
\[
a(t)+\beta(t)s = \Phi(t)s + \int_t^T \epsilon\Phi(u)A(u)du.
\]
Since this must hold for all s, by matching the coefficients we obtain:
\begin{align}
\notag\beta(t) = 1  \hspace{0.2cm}\forall t
\end{align}
as proven in \ref{stable_proof_1}. Now the solution for the deterministic part differs from the previous theorem for the performative stable point, as in the LHS of the equation we have the $a(t)$, whereas in the RHS we have the term $A(u)$. Specifically given now that $\beta(t) = 1$ it holds for $\theta_2(t)=-\frac{1}{2}\gamma \sigma^2(T-t)$.
We have now that:
\begin{align}
\notag a(t) &= \int_t^T \epsilon\Phi(u)A(u)du\xRightarrow{\frac{\partial}{\partial t}}\\
\notag a^{'}(t) &=-\epsilon\Phi(t)(a(t)+2q_t\theta_2(t)) \xRightarrow{b(t)=1\Rightarrow \Phi(t)=1}    \\
\notag a^{'}(t) &=-\epsilon\left(a(t)+2q_t(-\frac{1}{2}\gamma\sigma^2(T-t))\right) =-\epsilon\alpha(t)+\epsilon q_t \gamma \sigma^2(T-t)\\
\notag a^{'}(t) +\epsilon\alpha(t) &= \epsilon q_t \gamma \sigma^2(T-t) \xRightarrow{\text{Integrating factor: }e^{\epsilon t}}\\
\notag  \frac{\partial(a(t)e^{\epsilon t})}{\partial t} &= \epsilon q_t \gamma \sigma^2(T-t) e^{\epsilon t} \xRightarrow[\int_t^T]{a(T)=0}\\
\notag -a(t)e^{\epsilon t} &= \epsilon q_t \gamma \sigma^2\int_t^T (T-u) e^{\epsilon u}du.
\end{align}
The solution of the integral is:
\[
\int_t^T (T-u) e^{\epsilon u}du= \frac{e^{\epsilon t}\left( \epsilon(t-T)-1\right)+e^{\epsilon T}}{\epsilon^2}\,.
\]
Therefore, the above becomes
\begin{align}
\notag -a(t)e^{\epsilon t} &= \epsilon q_t \gamma \sigma^2 \left(   \frac{e^{\epsilon t}\left( \epsilon(t-T)-1\right)+e^{\epsilon T}}{\epsilon^2}        \right)           \\
\notag  a(t) &=  -\epsilon q_t \gamma \sigma^2 \left(   \frac{e^{\epsilon t}\left( \epsilon(t-T)-1\right)+e^{\epsilon T}}{\epsilon^2e^{\epsilon t}}\right) \Rightarrow          \\
\notag a(t)&= q_t\gamma\sigma^2(T-t)-q_t\gamma\sigma^2\frac{(e^{\epsilon(T-t)}-1)}{\epsilon}\,.
\end{align} 
Thus, we now have
\begin{align}
\notag& r^*_{PS}=a(t)+2q\theta_2(t)+\beta(t)s \Rightarrow\\
\notag& r^*_{PS}= q_t\gamma\sigma^2(T-t)-q_t\gamma\sigma^2\frac{(e^{\epsilon(T-t)}-1)}{\epsilon} -q_t(\gamma \sigma^2(T-t))+s\\
\notag& r^*_{PS}=s-q_t\gamma\sigma^2\frac{(e^{\epsilon(T-t)}-1)}{\epsilon}\,.
\end{align}


\subsubsection{Verification of the fixed point}

Conversely, any performatively stable reservation price must satisfy this fixed-point condition.
Hence, $r^*_{\mathrm{PS}}$ is a fixed point of the optimisation operator:
\[
r^*_{\mathrm{PS}} = \OptMM(r^*_{\mathrm{PS}}).
\]

To verify that with the analytical solution we obtained. Assume the reservation price $r^*=s_u-q_u\gamma\sigma^2\frac{(e^{\epsilon(T-u)}-1)}{\epsilon}$. Then the performative effect of that strategy will induce the SDE:
\begin{align*}
ds_u &= \epsilon\bigl(s_u-q_u\gamma\sigma^2\frac{(e^{\epsilon(T-u)}-1)}{\epsilon} - s_u\bigr)\,du + \sigma\,dW_u\\    
ds_u&= - q_t\gamma\sigma^2(e^{\epsilon(T-t)}-1)du + \sigma\,dW_u
\end{align*}
which yields:
\[
\mathbb{E}_{t,s}[S_T] =s_t - q_t\gamma\sigma^2\bigl(\frac{e^{\epsilon(T-t)}-1}{\epsilon}-(T-t)\bigr).
\]
We notice that the conditional expectation is affine to s with $\beta(t)=1$ and $\alpha(t)=- q_t\gamma\sigma^2\bigl(\frac{e^{\epsilon(T-t)}-1}{\epsilon}-(T-t\bigr)$. So, the general formulas of Theorem \ref{lem:generic_opt_solutions} apply and therefore:
\begin{align*}
\theta_1(t) &= \mathbb{E}_{t,s}[S_T]=s_t - q_t\gamma\sigma^2\bigl(\frac{e^{\epsilon(T-t)}-1}{\epsilon}-(T-t)\bigr)\\
\theta_2(t)&= - \frac{1}{2}\gamma\sigma^2 \int_t^T \beta^2(\xi,T)\,d\xi = - \frac{1}{2}\gamma\sigma^2\int_t^T 1\,d\xi=- \frac{1}{2}\gamma\sigma^2(T-t).
\end{align*}
For the optimal reservation price, it holds that
\begin{align*}
r^*_{PO} &= \theta_1(t,s)+2q\theta_2(t) = s_t - q_t\gamma\sigma^2\bigl(\frac{e^{\epsilon(T-t)}-1}{\epsilon}-(T-t)\bigr) +2q \left(-\frac{1}{2}\gamma\sigma^2(T-t)\right)\\
&=s_t - q_t\gamma\sigma^2(\frac{e^{\epsilon(T-t)}-1}{\epsilon})+ q_t\gamma\sigma^2(T-t)-q_t\gamma\sigma^2(T-t)=s_t - q_t\gamma\sigma^2(\frac{e^{\epsilon(T-t)}-1}{\epsilon})\\&=r^*
\end{align*}
which is the initial $r^*$ that we began with. So we have that $r^*_{PO}= \OptMM(r^*)=r^*$ Therefore $r^*$ is a performative stable strategy.

\section{Additional Simulation Results}
\begin{table*}[!htbp]
\centering
\caption{Performance metrics across all strategies for different \(\epsilon\) values; $\gamma=0.5$.}
\small
\resizebox{1.005\textwidth}{!}{
\begin{tabular}{@{}c@{\hspace{8pt}} c@{\hspace{3pt}}c@{\hspace{3pt}}r c@{\hspace{3pt}}c@{\hspace{3pt}}r
c@{\hspace{3pt}}c@{\hspace{3pt}}r@{}}
\toprule
& \multicolumn{3}{c@{\hspace{23pt}}}{A\&S Strat.} 
& \multicolumn{3}{c@{\hspace{23pt}}}{Symmetric Strat.} 
& \multicolumn{3}{c@{}}{Performative Strat.} \\
\cmidrule(lr){2-4} \cmidrule(lr){5-7} \cmidrule(lr){8-10}
\(\epsilon\) & Profit & Sharpe & \multicolumn{1}{r}{Term Inv.} 
            & Profit & Sharpe & \multicolumn{1}{r}{Term Inv.}
            & Profit & Sharpe & \multicolumn{1}{r@{}}{Term Inv.}\\
\midrule
0  & $52.87\pm5.43$ & $9.74$ & $0.06\pm2.52$
   & $59.69\pm9.57$ & $6.23$ & $0.50\pm8.20$
   & $52.93\pm5.35$ & \textbf{9.89} & $0.06\pm2.52$ \\
5  & $41.69\pm5.81$ & $7.18$ & $0.15\pm2.47$
   & $59.83\pm18.21$ & $3.28$ & $-0.03\pm7.85$
   & $47.74\pm13.35$ & $3.58$ & $0.14\pm3.61$ \\
10 & $30.71\pm7.91$ & $3.88$ & $-0.05\pm2.55$
   & $58.37\pm34.02$ & $1.72$ & $0.02\pm7.92$
   & \textbf{70.38$\pm$27.55} & $2.55$ & $0.05\pm5.35$ \\
15 & $19.84\pm10.76$ & $1.84$ & $0.07\pm2.47$
   & $61.45\pm46.18$ & $1.33$ & $0.33\pm7.68$
   & \textbf{90.70$\pm$39.80} & \textbf{2.28} & $0.24\pm6.13$ \\
19 & $11.28\pm12.19$ & $0.92$ & $-0.03\pm2.57$
   & $58.99\pm58.90$ & $1.00$ & $0.15\pm8.07$
   &\textbf{115.18$\pm$59.12} & \textbf{1.95} & $-0.38\pm7.51$ \\
\bottomrule
\end{tabular}
}
\label{tab:strategy_results}
\end{table*}

For every $\epsilon$, except $\epsilon=5$ either the total profit or the Sharpe Ratio of the performative aware strategy exceeds that of the symmetric strategy and A\&S strategy. Up to $\epsilon=10$ the performative aware strategy holds comparable with A\&S's terminal inventory both in mean and std values and always smaller terminal inventory statistics than that of symmetrical over either the mean or its variance.

\end{document}